\documentclass[12pt]{article}
\pdfoutput=1

\usepackage[pdftex]{graphicx}
\usepackage{amsmath,amssymb}
\usepackage{psfrag}
\usepackage[pdftex]{hyperref}

\newcounter{comment}

{\refstepcounter{comment}%
\begin{quote}
\ttfamily\small$\blacksquare$ \textbf{\underline{Comment} $\sharp$\thecomment:}}%
{\end{quote}}

{
\begin{quote}
\ttfamily\small$\blacktriangleright$ \textbf{\underline{Reply} $\sharp$\thecomment:}}%
{\end{quote}}

\font\cmss=cmss12 
\def\1{\hbox{{1}\kern-.25em\hbox{l}}}
\def\bfZ{\relax{\hbox{\cmss Z\kern-.4em Z}}}

\def\ru1{\rule[-0.4truecm]{0mm}{1truecm}}

\renewcommand{\thefootnote}{\fnsymbol{footnote}}
\renewcommand{\bar}[1]{\overline{#1}}

\def\ru1{\rule[-0.4truecm]{0mm}{1truecm}}
\def\upleftarrow#1{\overleftarrow{#1}}
\def\uprightarrow#1{\overrightarrow{#1}}
\def\thru#1{\mathrel{\mathop{#1\!\!\!/}}}
\def\be{\begin{eqnarray}}
\def\ee{\end{eqnarray}}
\def\bea{\begin{eqnarray}}
\def\eea{\end{eqnarray}}
\def\kT{{\bf k}_\perp}
\def\rT{{\bf r}_\perp}

\def\qT{{\bf q}_\perp}

\def\bT{{\bf b}_\perp}
\def\rT{{\bf r}_\perp}

\def\D2{{\bf \Delta}_\perp^2}
\def\0T{{\bf 0}_\perp}

\begin{document}



\setcounter{footnote}{0}
\renewcommand{\thefootnote}{\fnsymbol{footnote}}
\renewcommand{\bar}[1]{\overline{#1}}
\newcommand{\ie}{{\it i.e.}}
\newcommand{\eg}{{\it e.g.,}}
\newcommand{\btt}[1]{{\tt$\backslash$#1}}
\newcommand{\half}{{$\frac{1}{2}$}} 
\newcommand{\ket}[1]{\left\vert\,{#1}\right\rangle}
\newcommand{\VEV}[1]{\left\langle{#1}\right\rangle}

\def\ru1{\rule[-0.4truecm]{0mm}{1truecm}}
\def\upleftarrow#1{\overleftarrow{#1}}
\def\uprightarrow#1{\overrightarrow{#1}}
\def\thru#1{\mathrel{\mathop{#1\!\!\!/}}}

\def\senk#1{\bbox{#1}_\perp}
\def\ha{{1\over 2}}
\def\ub#1{\underline{#1}}
\def\ths{\thinspace}
\def\psibar{\overline{\psi}}
\def\del{\partial}
\def\ra{\rightarrow}
\def\eg{{\it e.g.}}
\def\g{\gamma}

\newpage
\begin{flushright}
May 2008
\end{flushright}

\bigskip

{\centerline{\Large \bf Charge Distributions in Transverse Coordinate Space}
{\centerline{\Large \bf and in Impact Parameter Space}}


\vspace{7mm}
\vspace{7mm}

\centerline{\bf Dae Sung Hwang$^{a}$, Dong Soo Kim$^{b}$, and Jonghyun Kim$^{a}$}

\vspace{3mm}

\vspace{4mm} \centerline{\it $^a$Department of Physics, Sejong University, Seoul
143--747, South Korea}
\vspace{1mm} \centerline{\it $^b$Department of Physics, Kangnung National University,
Kangnung 210-702, South Korea}

\vspace*{1.2cm}



\begin{abstract}
\noindent
We study the charge distributions of the valence quarks inside nucleon
in the transverse coordinate space,
which is conjugate to the transverse momentum space.
We compare the results with the charge distributions in the impact parameter space.
\end{abstract}

\vfill

\centerline{
}
\vfill


\newpage

\section{Introduction}
In recent years the role of the transverse momentum of the parton has
been more important in the field of the hadron physics since
it provides time-odd distribution and fragmentation functions,
and makes the single-spin asymmetries (SSA) in hadronic processes possible
\cite{MT96,BM98,BHS}.
One gluon exchange in the final state interactions (FSI) has been understood
as a mechanism for generating a transverse single-spin asymmetry \cite{BHS}.
This FSI can be effectively
taken into account by introducing an appropriate
Wilson line phase factor in the definition of the distribution
functions of quarks in the nucleon \cite{sivers,collins,ji1,ji2}.
It is possible when the distribution functions are functions of
the transverse momenta of the partons, as well as the longitudinal
momentum fractions.
Therefore, including the transverse momentum of the parton into consideration
enlarges the realm of the investigation of the nucleon structure.

In Ref. \cite{BHS} a simple scalar diquark model was used
to demonstrate explicitly that the FSI can indeed give rise to
a leading-twist transverse SSA, which emerged from interference
between spin dependent amplitudes with different 
nucleon spin states.
In Refs. \cite{BHS,weak} it was observed that the
same overlap integrals between light-cone wavefunctions
that describe the anomalous magnetic moment
also appear in the expression of the Sivers distribution function
with an additional factor in the integrand.
Since these integrals are the overlaps between light-cone wavefunctions
whose orbital angular momenta differ by $\Delta L^z = \pm 1$,
the orbital angular momentum of the quark inside the proton is
essential for the existence of the Sivers asymmetry.

In Refs. \cite{mb1,mb2} the single-spin asymmetries were analyzed in
the impact parameter space, in particular,
the transverse distortion of parton distributions \cite{burkardt00,burkardt03}
was used to develop a physical explanation for the sign of the SSAs
for transversely polarized targets.
These aspects were illustrated explicitly by using the scalar diquark model
in Ref. \cite{BH04}.
Miller obtained the charge distributions in the impact parameter space
for the valence quarks inside nucleon and found that the charge density
inside neutron is negative at the center \cite{miller07}.
The term `impact parameter space' in this paper means that defined in
these references.

In this paper we study the transverse coordinate space of the parton which
is the Fourier conjugate to the transverse momentum space.
We investigate the charge distributions in the transverse coordinate space
of the valence quarks inside proton and neutron, and compare them with
those in the impact parameter space \cite{miller07}.
We present the results in terms of the scalar diquark model \cite{BHS}
for simplicity of presentation, however, extending to more general systems is
straightforward.
We also show the difference of the charge distributions in the transverse
coordinate space and those in the impact parameter space clearly by using
an explicit example of scalar diquark model.

For explicit calculations we use the light-cone wavefunctions.
The light-cone (LC) Fock representation of composite systems such as hadrons in
QCD has a number of remarkable properties.
Because the generators of certain Lorentz boosts are kinematical, knowing
the wavefunction in one frame allows one to obtain it in any other frame
\cite{BL80,PinskyPauli}.
One can construct any electromagnetic,
electroweak, or gravitational form factor or local operator product
matrix element of a composite or elementary system from the
overlap of the LC wavefunctions
\cite{BD80,Brodsky:1998hn,BHMS}.
LC wavefunctions also provide a convenient representation of the generalized
parton distributions in terms of overlap integrals \cite{DFJK,BDH,HM07}.
In this paper we can study the charge distributions in the transverse coordinate
space efficiently in the light-cone framework.

\section{Charge Distributions in Transverse Coordinate Space}

We consider a scalar diquark model which is
an effective composite system composed of a fermion and
a neutral scalar based on the one-loop quantum fluctuations of
Yukawa theory.
The light-cone wavefuctions describe off-shell particles but are
computable explicitly from perturbation theory \cite{BHMS,HM07}.

The $J^z = + {1\over 2}$ two-particle Fock state is given by
\begin{eqnarray}
&&\left|\Psi^{\uparrow}_{\rm two \ particle}(P^+, \bf P_\perp = \bf
0_\perp)\right>
\label{sn1}\\
&=&
\int\frac{{\mathrm d}^2 {\bf k}_{\perp} {\mathrm d} x }{{\sqrt{x(1-x)}}16
\pi^3}
\Big[ \
\psi^{\uparrow}_{+\frac{1}{2}} (x,{\bf k}_{\perp})\,
\left| +\frac{1}{2}\, ;\,\, xP^+\, ,\,\, {\bf k}_{\perp} \right>
+\psi^{\uparrow}_{-\frac{1}{2}} (x,{\bf k}_{\perp})\,
\left| -\frac{1}{2}\, ;\,\, xP^+\, ,\,\, {\bf k}_{\perp} \right>\ \Big]\ ,
\nonumber
\end{eqnarray}
where
\begin{equation}
\left
\{ \begin{array}{l}
\psi^{\uparrow}_{+\frac{1}{2}} (x,{\bf k}_{\perp})=\frac{(xM+m)}{x}\,
\varphi(x,{\bf k}_\perp) \ ,\\
\psi^{\uparrow}_{-\frac{1}{2}} (x,{\bf k}_{\perp})=
-\frac{(+k^1+{\mathrm i} k^2)}{x }\,
\varphi(x,{\bf k}_\perp) \ ,
\end{array}
\right.
\label{sn2}
\end{equation}
in terms of a scalar function $\varphi(x,{\bf k}_\perp)$. This
scalar function arises from the spectator propagator in a triangle
Feynman diagram \cite{Brodsky:1998hn,BHMS} and so the underlying
Lorentz symmetry is respected. We generalize $\varphi(x,{\bf
k}_\perp)$ by an adjustment of its power behavior $p$:
\begin{eqnarray}
\label{Def-LC-WF3} \varphi(x,{\bf k}_\perp) = \frac{g
M^{2p}}{\sqrt{1-x}} x^{-p} \left( M^2 - \frac{{\bf
k}_\perp^2+m^2}{x} -\frac{{\bf k}_\perp^2+\lambda^2}{1-x}
\right)^{-p-1}\,,
\end{eqnarray}
where $M$, $\lambda$ and $m$ are the proton, spectator, and quark
masses, respectively.  The  Yukawa theory result is for $p=0$, and
Eq. (\ref{Def-LC-WF3}) has an additional factor
$M^{2p}x^{-p}\left( M^2 - \frac{{\bf k}_\perp^2+m^2}{x}
-\frac{{\bf k}_\perp^2+\lambda^2}{1-x} \right)^{-p}$ compared to
the scalar function for the Yukawa model presented in Ref.
\cite{BHMS}. In this additional factor, $M^{2p}$ is
attached for the dimensional purpose and the remaining factor can
be induced from a Lorentz invariant form factor $(k^2-m^2)^{-p}$
at the quark-diquark vertex as in Ref.~\cite{mulders97}.

Similarly, the $J^z = - {1\over 2}$ two-particle Fock state is given by
\begin{eqnarray}
&&\left|\Psi^{\downarrow}_{\rm two \ particle}(P^+, \bf P_\perp =
\bf 0_\perp)\right>
\label{sn1a}\\
&=&
\int\frac{{\mathrm d}^2 {\bf k}_{\perp} {\mathrm d} x }{{\sqrt{x(1-x)}}16
\pi^3}
\Big[ \
\psi^{\downarrow}_{+\frac{1}{2}} (x,{\bf k}_{\perp})\,
\left| +\frac{1}{2}\, ;\,\, xP^+\, ,\,\, {\bf k}_{\perp} \right>
+\psi^{\downarrow}_{-\frac{1}{2}} (x,{\bf k}_{\perp})\,
\left| -\frac{1}{2}\, ;\,\, xP^+\, ,\,\, {\bf k}_{\perp} \right>\ \Big]\ ,
\nonumber
\end{eqnarray}
where
\begin{equation}
\left
\{ \begin{array}{l}
\psi^{\downarrow}_{+\frac{1}{2}} (x,{\bf k}_{\perp})=
\frac{(+k^1-{\mathrm i} k^2)}{x }\,
\varphi(x,{\bf k}_\perp) \ ,\\
\psi^{\downarrow}_{-\frac{1}{2}} (x,{\bf k}_{\perp})=\frac{(xM+m)}{x}\,
\varphi(x,{\bf k}_\perp) \ .
\end{array}
\right.
\label{sn2a}
\end{equation}
In (\ref{sn2}) and (\ref{sn2a}) we have generalized the framework of
the Yukawa theory by assigning a mass $M$ to the external electrons,
but a different mass $m$ to the internal quark (fermion)
line and a mass $\lambda$ to the internal diquark (boson) line
\cite{BD80}.  The idea behind this is to model the structure
of a composite fermion state with mass $M$ by a fermion and a boson
constituent with respective masses $m$ and $\lambda$.

The LC wavefunction in the transverse coordinate space $\tilde{\psi}(x,\rT)$
is given by the Fourier transformation of $\psi(x,\kT)$:
\begin{eqnarray}
\psi(x,\kT)&=& \int {d^2\rT}
{\rm exp}\Big[ {-i{{\bf k}_\perp}\cdot{{\bf r}_\perp}}\Big]
\tilde{\psi}(x,\rT) \nonumber\\
\tilde{\psi}(x,\rT)&=& \int \frac{d^2 \kT}{(2\pi)^2}
{\rm exp}\Big[ {i{{\bf k}_\perp}\cdot{{\bf r}_\perp}}\Big]
{\psi}(x,\kT)\ .
\label{coorwf2}
\end{eqnarray}
From (\ref{coorwf2}) we have the relation
\begin{equation}
\int \frac{d^2 \kT}{(2\pi)^2}{\psi}^*(x,\kT){\psi}(x,\kT)=
\int {d^2\rT}\tilde{\psi}^*(x,\rT)\tilde{\psi}(x,\rT)\ .
\label{momcoor2}
\end{equation}

(\ref{sn2}), (\ref{sn2a}) and (\ref{coorwf2}) give the following light-cone
wavefunctions in the transverse coordinate space:
\begin{equation}
\left
\{ \begin{array}{l}
\tilde{\psi}^{\uparrow}_{+\frac{1}{2}} (x,{\bf r}_{\perp})=
(xM+m){r_{\perp}\over A_R}{gM^2\over 4\pi}(1-x)^{3\over 2}K_1(r_{\perp}A_R)
\ ,\\
\tilde{\psi}^{\uparrow}_{-\frac{1}{2}} (x,{\bf r}_{\perp})=
-{\mathrm i} (r^1+{\mathrm i} r^2)
{gM^2\over 4\pi}(1-x)^{3\over 2}K_0(r_{\perp}A_R)
\ ,
\end{array}
\right.
\label{sn2rp}
\end{equation}
and
\begin{equation}
\left
\{ \begin{array}{l}
\tilde{\psi}^{\downarrow}_{+\frac{1}{2}} (x,{\bf r}_{\perp})=
{\mathrm i} (r^1-{\mathrm i} r^2)
{gM^2\over 4\pi}(1-x)^{3\over 2}K_0(r_{\perp}A_R)
\ ,\\
\tilde{\psi}^{\downarrow}_{-\frac{1}{2}} (x,{\bf r}_{\perp})=
(xM+m){r_{\perp}\over A_R}{gM^2\over 4\pi}(1-x)^{3\over 2}K_1(r_{\perp}A_R)
\ ,
\end{array}
\right.
\label{sn2arp}
\end{equation}
where ${\bf r}_{\perp}=(r^1,r^2)$, $r_{\perp}=|{\bf r}_{\perp}|$ and
\begin{equation}
A_R^2=-M^2x(1-x)+m^2(1-x)+{\lambda_R}^2x, \ \ \ \
R=u\ {\rm or}\ d\ .
\label{ar12}
\end{equation}
Then, in the scalar diquark model
the distribution in the transverse momentum space and that in the
transverse coordinate space are given, respectively, by
\begin{eqnarray}
P(\kT)&=&
\int {\mathrm d} x\, P(x,\kT)\ =\
\int \frac{{\mathrm d} x}{(2\pi)^2}\ 
\Big[\psi^{\uparrow\ *}_{+\frac{1}{2}}(x,{\bf k}_{\perp})
\psi^{\uparrow}_{+\frac{1}{2}}(x,{\bf k}_{\perp})
+\psi^{\uparrow\ *}_{-\frac{1}{2}}(x,{\bf k}_{\perp})
\psi^{\uparrow}_{-\frac{1}{2}}(x,{\bf k}_{\perp})\Big]
\ ,\ \ \
\label{diskr1}\\
P(\rT)&=&
\int {\mathrm d} x\, P(x,\rT)\ =\
\int {\mathrm d} x\ 
\Big[\tilde{\psi}^{\uparrow\ *}_{+\frac{1}{2}}(x,\rT)
\tilde{\psi}^{\uparrow}_{+\frac{1}{2}}(x,\rT)
+\tilde{\psi}^{\uparrow\ *}_{-\frac{1}{2}}(x,\rT)
\tilde{\psi}^{\uparrow}_{-\frac{1}{2}}(x,\rT)\Big]\ .
\label{diskr2}
\end{eqnarray}

\section{Comparison of Charge Distributions in Transverse Coordinate and
Impact Parameter Spaces}

Miller calculated the parton charge densities of nucleons by using the formula
\cite{miller07}
\begin{equation}
\rho^{\rm Imp}({\bf b}_\perp)=
\int \frac{d^2 \qT}{(2\pi)^2}
{\rm exp}\Big[ {i{{\bf q}_\perp}\cdot{{\bf b}_\perp}}\Big]
F_1(\qT)\ .
\label{gmimp}
\end{equation}
In this section we interpret this $\rho^{\rm Imp}({\bf b}_\perp)$ in terms of
the LC wavefunctions and compare it with the charge density in the transverse
coordinate space $P(\rT)$ given in (\ref{diskr2}).

The fact that certain amplitudes that are convolutions in
momentum space become diagonal in position space can be
easily understood on the basis of some elementary theorems
about convolutions and Fourier transforms. For example, if
\bea
f(\kT)&=& \int {d^2\rT} {\rm exp}\Big[ {-i{{\bf k}_\perp}\cdot{{\bf r}_\perp}}\Big]
\tilde{f}(\rT) \nonumber\\
g(\kT)&=&\int {d^2\rT} {\rm exp}\Big[ {-i{{\bf k}_\perp}\cdot{{\bf r}_\perp}}\Big]
\tilde{g}(\rT)
\eea
then the ``form factor''
\bea
F(\qT)\equiv \int \frac{d^2\kT}{(2\pi)^2} f^*(\kT-\qT)g(\kT)
\eea
becomes diagonal in Fourier space
\bea
\int \frac{d^2 \qT}{(2\pi)^2}
{\rm exp}\Big[ {i{{\bf q}_\perp}\cdot{{\bf r}_\perp}}\Big] F(\qT) =
\tilde{f}^*(\rT) \tilde{g}(\rT).
\eea
This well-known result forms the basis for the interpretation
of non-relativistic form factors as charge distributions
in position space.

On the other hand, for
\bea
G(\qT)\equiv \int \frac{d^2\kT}{(2\pi)^2} f^*(\kT-a\qT)g(\kT)
\eea
we have
\bea
\int \frac{d^2 \qT}{(2\pi)^2}
{\rm exp}\Big[ {i{{\bf q}_\perp}\cdot{{\bf b}_\perp}}\Big] G(\qT) =
{1\over |a|^2}\ \tilde{f}^*({\bT\over a}) \tilde{g}({\bT\over a}).
\eea

$F_1(\qT)$ can be expressed as
\begin{eqnarray}
F_1(\qT)&=&\int{\mathrm d} x\ H(x,0,\qT)
\nonumber\\
&=&
\int\frac{{\mathrm d}^2 {\bf k}_{\perp} {\mathrm d} x }{(2 \pi)^2}
\Big[\psi^{\uparrow\ *}_{+\frac{1}{2}}(x,{\bf k'}_{\perp})
\psi^{\uparrow}_{+\frac{1}{2}}(x,{\bf k}_{\perp})
+\psi^{\uparrow\ *}_{-\frac{1}{2}}(x,{\bf k'}_{\perp})
\psi^{\uparrow}_{-\frac{1}{2}}(x,{\bf k}_{\perp})\Big]\ ,
\label{BDF1as}
\end{eqnarray}
where $P'=P+q$ and
\begin{equation}
{\bf k'}_{\perp}={\bf k}_{\perp}+(1-x){\bf q}_{\perp}\ .
\label{BDF1b}
\end{equation}
We note that in (\ref{BDF1as}) we adopted the normalization of the wavefunction,
with which $P(\kT)$ in (\ref{diskr1}) satisfies
$F_1(0)=\int{\mathrm d}^2 {\bf k}_{\perp} P(\kT)=1$.

Then, for $\rho^{\rm Imp}({\bf b}_\perp)$ given in (\ref{gmimp}), we have
\begin{eqnarray}
&&\rho^{\rm Imp}({\bf b}_\perp)\ =\
\int {\mathrm d} x\, \rho^{\rm Imp}(x,{\bf b}_\perp)
\nonumber\\ 
&=&
\int \frac{d^2 \qT}{(2\pi)^2}
{\rm exp}\Big[ {i{{\bf q}_\perp}\cdot{{\bf b}_\perp}}\Big]
F_1(\qT)
\ =\
\int \frac{d^2 \qT}{(2\pi)^2}
{\rm exp}\Big[ {i{{\bf q}_\perp}\cdot{{\bf b}_\perp}}\Big]
\int{\mathrm d} x\ H(x,0,\qT)
\nonumber\\ 
&=&\int{\mathrm d} x \ {1\over (1-x)^2}
\Big[\tilde{\psi}^{\uparrow\ *}_{+\frac{1}{2}}(x,{\bT\over -1+x})
\tilde{\psi}^{\uparrow}_{+\frac{1}{2}}(x,{\bT\over -1+x})
+\tilde{\psi}^{\uparrow\ *}_{-\frac{1}{2}}(x,{\bT\over -1+x})
\tilde{\psi}^{\uparrow}_{-\frac{1}{2}}(x,{\bT\over -1+x})\Big]
\ ,\ \
\label{ffaa}
\end{eqnarray}
where $\tilde{\psi}(x,\rT)$ is the Fourier conjugate to $\psi(x,\kT)$
as defined in (\ref{coorwf2}).
From (\ref{ffaa}) we have
\begin{equation}
\rho^{\rm Imp}(x,{\bf b}_\perp)=
\int \frac{d^2 \qT}{(2\pi)^2}
{\rm exp}\Big[ {i{{\bf q}_\perp}\cdot{{\bf b}_\perp}}\Big]
H(x,0,\qT)
={1\over (1-x)^2}\, P(x,{\bT\over -1+x})\ ,
\label{comp1}
\end{equation}
where $P(x,\rT)$ in transverse coordinate space is given in (\ref{diskr2}) as
\begin{equation}
P(x,\rT)=\Big[\tilde{\psi}^{\uparrow\ *}_{+\frac{1}{2}}(x,\rT)
\tilde{\psi}^{\uparrow}_{+\frac{1}{2}}(x,\rT)
+\tilde{\psi}^{\uparrow\ *}_{-\frac{1}{2}}(x,\rT)
\tilde{\psi}^{\uparrow}_{-\frac{1}{2}}(x,\rT)\Big]\ .
\label{diskr2aa}
\end{equation}
Eq. (\ref{comp1}) shows clearly the relation between the parton charge
density in the impact parameter $(x,{{\bf b}_\perp})$ space
$\rho^{\rm Imp}(x,{\bf b}_\perp)$
and the distribution in the transverse coordinate $(x,{{\bf r}_\perp})$ space
$P(x,{{\bf r}_\perp})$.

In this paper $P(x,\rT)$ and $\rho^{\rm Imp}(x,{\bf b}_\perp)$ denote
parton density distributions for $u$ and $d$ quarks inside proton,
and charge distributions for proton ($p$) and neutron ($n$).
That is, $P_p={4\over 3}P_u-{1\over 3}P_d$ and $P_n={2\over 3}P_d-{2\over 3}P_u$
from the isospin symmetry,
and the same relations for $\rho^{\rm Imp}(x,{\bf b}_\perp)$'s.

\section{Explicit Example with Scalar Diquark Model}

\begin{figure}
\centering
\psfrag{xperp}[cc][cc]{$x_\perp$}
\begin{minipage}[t]{8.0cm}
\centering
\includegraphics[width=\textwidth]{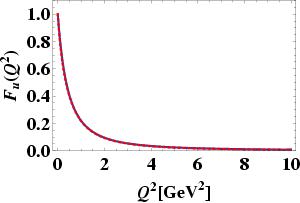}
(a)
\end{minipage}\hspace{1.0cm}
\begin{minipage}[t]{8.0cm}
\centering
\includegraphics[width=\textwidth]{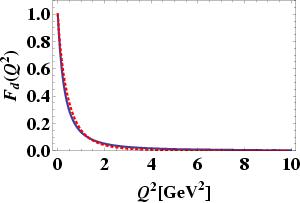}
(b)
\end{minipage}
\parbox{0.95\textwidth}{\caption{
The fitting of the Dirac form factors of $u$ and $d$ quarks in proton
by using  the scalar diquark model with
$m=0.5$ GeV, $\lambda_u = 0.63$ GeV and $\lambda_d = 0.535$ GeV.
The dotted lines are experimental results parameterized in \cite{kelly04}
and the solid lines are those from (\ref{BDF1as}) with the fitted parameter
values.
\label{fitF1u50}}}
\end{figure}

\begin{figure}
\centering
\psfrag{xperp}[cc][cc]{$x_\perp$}
\begin{minipage}[t]{8.0cm}
\centering
\includegraphics[width=\textwidth]{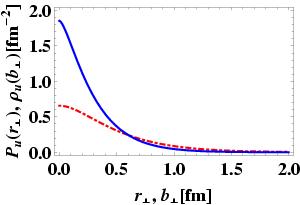}
(a)
\end{minipage}\hspace{1.0cm}
\begin{minipage}[t]{8.0cm}
\centering
\includegraphics[width=\textwidth]{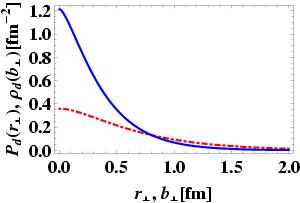}
(b)
\end{minipage}
\parbox{0.95\textwidth}{\caption{
The results of $P(\rT)$ (dash-dot line) and $\rho^{\rm Imp}(\bT)$ (solid line)
for $u$ and $d$ quarks inside proton for the fitted scalar diquark model.
\label{uwf50}}}
\end{figure}

\begin{figure}
\centering
\psfrag{xperp}[cc][cc]{$x_\perp$}
\begin{minipage}[t]{8.0cm}
\centering
\includegraphics[width=\textwidth]{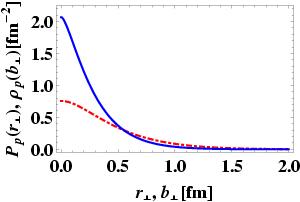}
(a)
\end{minipage}\hspace{1.0cm}
\begin{minipage}[t]{8.0cm}
\centering
\includegraphics[width=\textwidth]{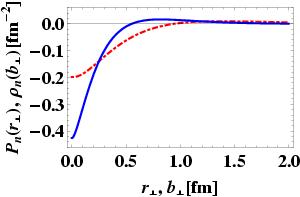}
(b)
\end{minipage}
\parbox{0.95\textwidth}{\caption{
The results of $P(\rT)$ (dash-dot line) and $\rho^{\rm Imp}(\bT)$ (solid line)
for proton and neutron for the fitted scalar diquark model.
\label{pxw50}}}
\end{figure}

Using the scalar diquark model given in (\ref{sn2}), (\ref{Def-LC-WF3}) and
(\ref{sn2a}), we fit the parameterization of Ref. \cite{kelly04} for
the experimental results of the Dirac form factors $F_1(q^2)$ of nucleons.
The parameterization of \cite{bradford06} is very similar to \cite{kelly04}.
We use $p=1$ in (\ref{Def-LC-WF3}).
The fitting of the Dirac form factors of $u$ and $d$ quarks in proton with
$m=0.5$ GeV, $\lambda_u = 0.63$ GeV and $\lambda_d = 0.535$ GeV are shown
in Fig. \ref{fitF1u50},
where the dotted lines are experimental results parameterized in \cite{kelly04}
and the solid lines are those from (\ref{BDF1as}) with these fitted values of
$m$, $\lambda_u$ and $\lambda_d$.

Fig. \ref{uwf50} presents the results of $P(\rT)$ and $\rho^{\rm Imp}(\bT)$
for $u$ and $d$ quarks inside proton, which are obtained by
using the LC wavefunctions (\ref{sn2}), (\ref{sn2a}), (\ref{sn2rp}) and (\ref{sn2arp}).
The dash-dot lines are graphs of $P(\rT)$ from (\ref{diskr2})
and the solid lines are $\rho^{\rm Imp}(\bT)$ obtained
from the formula (\ref{gmimp}) with $F_1(\qT)$ given by using (\ref{BDF1as}).
In the figures $\rho$ represents $\rho^{\rm Imp}$.
Fig. \ref{pxw50} presents the charge distributions inside proton and neutron
from the results of Fig. \ref{uwf50} by using $P_u=P_p+P_n/2$, $P_d=P_p+2P_n$
from the isospin symmetry, and the same relations for $\rho^{\rm Imp}(x,{\bf b}_\perp)$'s.

Figs. \ref{uwf50} and \ref{pxw50} show that $P(\rT)$ extends toward outside further than
$\rho^{\rm Imp}(\bT)$, which can be understood from the fact that $\bT / (-1+x)$
appears in the place of $\rT$ in Eqs. (\ref{ffaa}) and (\ref{comp1}).
This property is also shown clearly in Table 1 which presents the results of the average
values of $|\rT |$ and $|\bT |$.

\begingroup
\begin{table}[t]
\vspace*{1.2cm}
\label{tab:eigenvalue}
\begin{center}
\begin{tabular}{|c|c|c|c|c|}
\hline\hline
 &$u$&$d$&$p$&$n$\\
\hline\hline
\ \ \ \ \ \ $<r_{\perp}>$ \ \ \ \ \ \ & \ \ \ \ \ 0.77 \ \ \ \ \ &
\ \ \ \ \ 1.08 \ \ \ \ \  & \ \ \ \ \ 0.67 \ \ \ \ \ & \ \ \ \ \ 0.21 \ \ \ \ \  \\
\hline
\ \ \ \ \ \ $<b_{\perp}>$ \ \ \ \ \ \ & \ \ \ \ \ 0.50 \ \ \ \ \ &
\ \ \ \ \ 0.61 \ \ \ \ \  & \ \ \ \ \ 0.46 \ \ \ \ \  & \ \ \ \ \ 0.07 \ \ \ \ \ \\
\hline\hline
\end{tabular}
\end{center}
\vspace*{-0.5cm}
\caption{The average values of $|\rT |$ and $|\bT |$
for the fitted scalar diquark model.}
\end{table}
\endgroup

We could see the difference between $P(\rT)$ and
$\rho^{\rm Imp}(\bT)$ explicitly in Figs. \ref{uwf50}
and \ref{pxw50}, and in Table 1.
Furthermore, it should be useful to analyze both $x$ and $\rT / \bT$ dependences of
this property by comparing
$P(x,\rT)$ in (\ref{diskr2}) and $\rho^{\rm Imp}(x,{\bf b}_\perp)$ in (\ref{ffaa}).
Figs. \ref{puxr}, \ref{pdxr}, \ref{ppxr} and \ref{pnxr} present
$P(x,\rT)$ and $\rho^{\rm Imp}(x,{\bf b}_\perp)$ of
$u$ and $d$ quarks inside proton, and proton and neutron, respectively.
We can see in these figures that $P(x,\rT)$ decreases more slowly than
$\rho^{\rm Imp}(x,{\bf b}_\perp)$ when $\rT$ and ${\bf b}_\perp$ increase.
In order to show this property more clearly, we present in Fig. \ref{prx35}
their differences for a fixed value $x=0.35$.

\begin{figure}
\centering
\psfrag{xperp}[cc][cc]{$x_\perp$}
\begin{minipage}[t]{8.0cm}
\centering
\includegraphics[width=\textwidth]{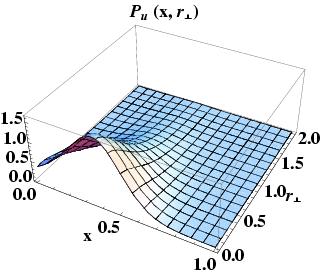}
(a)
\end{minipage}\hspace{1.0cm}
\begin{minipage}[t]{8.0cm}
\centering
\includegraphics[width=\textwidth]{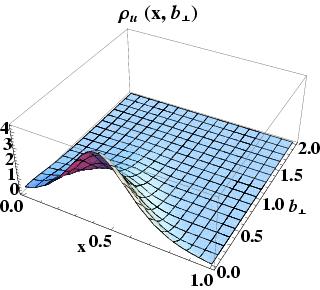}
(b)
\end{minipage}
\parbox{0.95\textwidth}{\caption{
$P(x,\rT)$ and $\rho^{\rm Imp}(x,{\bf b}_\perp)$ of
the $u$ quark inside proton
for the fitted scalar diquark model.
\label{puxr}}}
\end{figure}

\begin{figure}
\centering
\psfrag{xperp}[cc][cc]{$x_\perp$}
\begin{minipage}[t]{8.0cm}
\centering
\includegraphics[width=\textwidth]{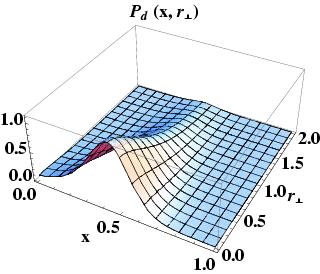}
(a)
\end{minipage}\hspace{1.0cm}
\begin{minipage}[t]{8.0cm}
\centering
\includegraphics[width=\textwidth]{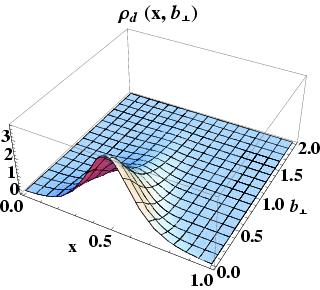}
(b)
\end{minipage}
\parbox{0.95\textwidth}{\caption{
$P(x,\rT)$ and $\rho^{\rm Imp}(x,{\bf b}_\perp)$ of
the $d$ quark inside proton
for the fitted scalar diquark model.
\label{pdxr}}}
\end{figure}

\begin{figure}
\centering
\psfrag{xperp}[cc][cc]{$x_\perp$}
\begin{minipage}[t]{8.0cm}
\centering
\includegraphics[width=\textwidth]{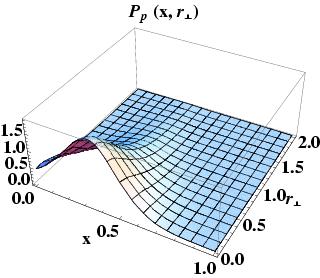}
(a)
\end{minipage}\hspace{1.0cm}
\begin{minipage}[t]{8.0cm}
\centering
\includegraphics[width=\textwidth]{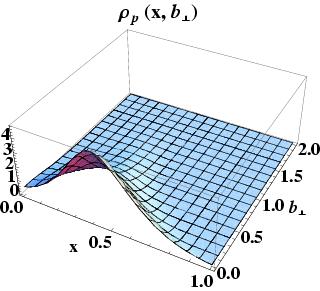}
(b)
\end{minipage}
\parbox{0.95\textwidth}{\caption{
$P(x,\rT)$ and $\rho^{\rm Imp}(x,{\bf b}_\perp)$ of
the proton
for the fitted scalar diquark model.
\label{ppxr}}}
\end{figure}

\begin{figure}
\centering
\psfrag{xperp}[cc][cc]{$x_\perp$}
\begin{minipage}[t]{8.0cm}
\centering
\includegraphics[width=\textwidth]{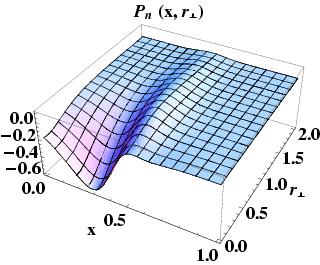}
(a)
\end{minipage}\hspace{1.0cm}
\begin{minipage}[t]{8.0cm}
\centering
\includegraphics[width=\textwidth]{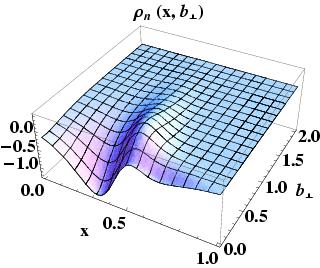}
(b)
\end{minipage}
\parbox{0.95\textwidth}{\caption{
$P(x,\rT)$ and $\rho^{\rm Imp}(x,{\bf b}_\perp)$ of
the neutron
for the fitted scalar diquark model.
\label{pnxr}}}
\end{figure}

\begin{figure}
\centering
\psfrag{xperp}[cc][cc]{$x_\perp$}
\begin{minipage}[t]{8.0cm}
\centering
\includegraphics[width=\textwidth]{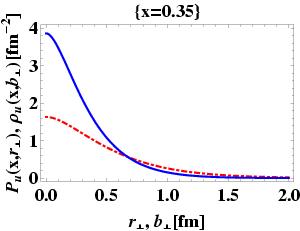}
(a)
\end{minipage}\hspace{1.0cm}
\begin{minipage}[t]{8.0cm}
\centering
\includegraphics[width=\textwidth]{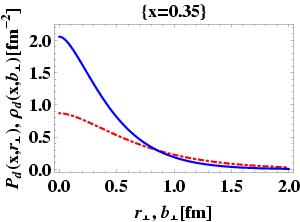}
(b)
\end{minipage}
\begin{minipage}[t]{8.0cm}
\centering
\includegraphics[width=\textwidth]{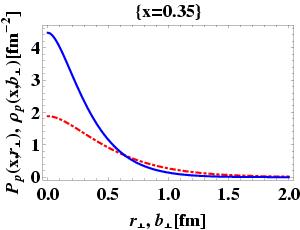}
(c)
\end{minipage}\hspace{1.0cm}
\begin{minipage}[t]{8.0cm}
\centering
\includegraphics[width=\textwidth]{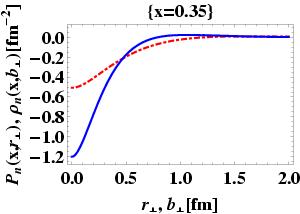}
(d)
\end{minipage}
\parbox{0.95\textwidth}{\caption{
$P(x=0.35,\rT)$ and $\rho^{\rm Imp}(x=0.35,{\bf b}_\perp)$ of
the $u$ quark inside proton $(a)$, the $d$ quark inside proton $(b)$,
the proton (c), and the neutron (d)
for the fitted scalar diquark model.
\label{prx35}}}
\end{figure}



\section{Conclusion}

The transverse momentum dependent distribution functions provide detailed
information on the nucleon structure. Then,
it is natural to investigate at the same time 
the distribution functions also in the transverse coordinate space, in order to
obtain the knowledge on the spatial structure of the nucleon.
For this purpose there have been a lot of interesting studies on distribution
functions in the impact parameter space, in particular in connection with the
generalized parton distribution functions, and there have been many important progresses.
However, it is desirable to understand better the relation of the impact parameter space
to the transverse (spatial) coordinate space.
In this paper we showed explicitly in the scalar diquark model
the relation between the charge distributions
inside nucleon in the transverse coordinate space $P(x,\rT)$
and those in the impact parameter space $\rho^{\rm Imp}(x,{\bf b}_\perp)$.
The figures of the results show that $P(x,\rT)$ decreases more slowly than
$\rho^{\rm Imp}(x,{\bf b}_\perp)$ when $\rT$ and ${\bf b}_\perp$ increase.
This property can be understood from the fact that ${\bT \over (-1+x)}$
appears in the argument of $P(x,\rT)$ in the formula given in (\ref{comp1}):
$\rho^{\rm Imp}(x,{\bf b}_\perp)={1\over (1-x)^2}\, P(x,{\bT\over -1+x})$.
As a consequence, $P(\rT)$ of (\ref{diskr2}) extends toward outside further than
$\rho^{\rm Imp}(\bT)$ of (\ref{gmimp}).
The results in this paper are also useful for the improvement of
understanding the relation between the transverse coordinate space and the impact
parameter space in general.

\section*{Acknowledgments}
This work was supported in part by the International Cooperation
Program of the KICOS (Korea Foundation for International Cooperation
of Science \& Technology),
in part by the 2007 research fund from Kangnung National University,
and in part by Seoul Fellowship.

\vspace*{1.0cm}


\end{document}